\documentclass{skads2009}
\usepackage{graphicx}

\usepackage{amsmath}
\usepackage{amssymb}
\usepackage{eurosym}

\usepackage{txfonts}
\usepackage{url}

\usepackage{longtable}
\usepackage{multirow}

\usepackage{subfigure}
\usepackage{epsfig}

\usepackage{natbib}
\usepackage{chapterbib}
\bibpunct{(}{)}{;}{a}{}{,}
\usepackage{rotating}

\newcommand{\etal}{et~al.}

\newcounter{chapter}

\newcommand{\mktitlecontents}[1]{
\addcontentsline{toc}{chapter}{{#1}\dotfill}
}
\newcommand{\mkauthorcontents}[1]{
\addtocontents{toc}{\newline\hspace*{1em}{\it\hangindent=1em {#1}}\\[0.5ex]}
}

\newcommand{\Title}[1]{
\title{#1}
\mktitlecontents{#1}
}
\newcommand{\Author}[1]{
\author{#1}
\mkauthorcontents{#1}
}

\newcommand{\Authorrunning}[1]{\authorrunning{#1}}
\newcommand{\Titlerunning}[1]{\titlerunning{#1}}

\begin{document}
\renewcommand{\thepage}{\arabic{page}}
\twocolumn

\Title{Probing the Epoch of Reionization with Low Frequency Arrays}
\Titlerunning{Probing the Epoch of Reionization with Low Frequency Arrays}

\Author{S.~Zaroubi}
\Authorrunning{S.~Zaroubi}
\institute{Kapteyn Astronomical Institute,
University of Groningen, Landleven 12, 9747AD Groningen, The
Netherlands}

   \abstract{The Epoch of Reionization (EoR) is the epoch in which
hydrogen in the Universe reionize after the ``Dark Ages''. This is the
second of two major phase transitions that hydrogen in the Universe
underwent, the first phase being the recombination era in which
hydrogen became neutral at redshift $\approx1100$. The EoR, occurs
around $z\approx 10$ and is probably caused by the first radiation
emitting astrophysical sources, hence it is crucial to our
understanding of when and how the Universe ``decided'' to start
forming astrophysical objects and how that influenced subsequent
structure formation in the Universe. As such, the EoR is related to
many fundamental questions in cosmology, galaxy formation, quasars and
very metal poor stars; all are foremost research issues in modern
astrophysics. The redshifted 21~cm hyperfine line is widely considered
as the most promising probe for studying the EoR in detail. In the
near future a number of low frequency radio telescopes (LOFAR, MWA,
GMRT and SKA) will be able to observe the 21~cm radiation arriving
from the high redshift Universe.  In this paper I present our current
picture of the ionization process, review the 21~cm line physics and
discuss the challenges that the current generation experiments are
expected to face. Finally, I discuss the potential of SKA in exploring
the EoR and the Universe's \textit{Dark Ages}.  }

\newcommand{\sn}{\smallskip\noindent} \def\dnsig{$D_n-\sigma$}
\def\ltsima{$\; \buildrel < \over \sim \;$}
\def\lsim{\lower.5ex\hbox{\ltsima}} \def\gtsima{$\; \buildrel > \over\sim \;$} 
\def\gsim{\lower.5ex\hbox{\gtsima}}
\def\ga{\mathrel{\hbox{\rlap{\hbox{\lower4pt\hbox{$\sim$}}}\hbox{$>$}}}}

  \maketitle
\label{zaroubi}
%
%

\section{Introduction}

\sn The last couple of decades have witnessed the emergence of an
overarching paradigm, the $\Lambda$CDM model, that describes the
formation and evolution of the Universe and its structure.  The
$\Lambda$CDM model accounts very successfully for most of the
available observational evidence on large scales. According to this
paradigm at redshift of 1100, about 400,000 years after the Big Bang,
the Universe's temperature and density decreased enough to allow the
ions and electrons to recombine into neutral hydrogen and helium (and
a very small percentage of heavier elements). Immediately afterwords,
photons decoupled from the baryonic material and the Universe became
transparent leaving a relic radiation, known as the cosmic microwave
background (CMB) radiation. To date, the CMB radiation has been
observed in many experiments providing one of the most compelling
evidences for the CDM paradigm \citep[for recent results see the WMAP
papers, e.g.,][]{spergel07,page07,komatsu10}.

\sn The matter-radiation decoupling has ushered the Universe into a
period of \textit{darkness} as its temperature dropped below 3000 K
and steadily decreased with the Universe's expansion. These
\textit{Dark Ages} ended about 400 million years later, when the first
radiation emitting objects (stars, black-holes, etc.)  were formed and
assembled into protogalaxies. The first objects then began to produce
a spectrum of radiation, especially ionizing ultra-violet photons.
After a sufficient number of ionizing sources had formed, the
temperature and the ionized fraction of the gas in the Universe
increased rapidly and most of the neutral hydrogen eventually ionized.
This period, during which the cosmic gas went from being almost
completely neutral to almost completely ionized, is known as the
\textit{Epoch of Reionization} of the Universe.

\sn The most accepted picture on how reionization unfolds is
simple. The first radiation-emitting objects ionize their immediate
surroundings, forming ionized bubbles that expand until the neutral
intergalactic medium (hereafter, IGM) consumes all ionizing
photons. As the number of objects increases, so do the number and size
of the bubbles, until they eventually fill the whole
Universe. However, most of the details of this scenario are yet to be
clarified.  For example: what controls the formation of the first
objects and how much ionizing radiation do they produce? How do the
bubbles expand into the intergalactic medium and what do they ionize
first, high-density or low density regions? The answer to these
questions and many others that arise in the context of studying the
EoR touches upon many fundamental questions in cosmology, galaxy
formation, quasars and very metal poor stars; all are foremost
research issues in modern astrophysics. A substantial theoretical
effort is currently dedicated to understanding the physical processes
that trigger this epoch, govern its evolution, and the ramifications
it had on subsequent structure formation \citep[c.f.,][]{barkana01,
bromm04, ciardi05, choudhury06, furlanetto06}. However, and despite
the pivotal role played by the EoR in cosmic history, observational
evidence on it is very scarce and, when available, is indirect and
model dependent.

\sn It is generally acknowledged that the 21cm emission line from
neutral hydrogen at high redshifts is the most promising probe for
studying the \textit{Dark Ages} and the EoR in detail. HI fills the
IGM except in regions surrounding the ionizing radiation of the first
objects to condense out of the cosmic flow.  Computer simulations
suggest that we may expect an evolving complex patch work of neutral
(HI) and ionized hydrogen (HII) regions
\citep{gnedin01,ciardi01,ritzerveld03,susa06,razoumov05,
nakamoto01,whalen06,rijkhorst05,mellema06,zahn07,mesinger07,pawlik08,
thomas09}.  The current constraint strongly suggest that the EoR
roughly straddles the redshift range of $z \sim20-6$.

\sn In this contribution I will give a short overview of the present
picture of the Universe's reionization and review the potential of
current and future instruments --focusing mostly on LOFAR and SKA-- in
detecting it.  The paper is organized as follows. In
section~\ref{sec:constraints} a review of the current observational
constraint is given. In section~\ref{sec:sources} the possible
reionization sources are discussed. Section~\ref{sec:signal} presents
the physics of the 21~cm line with the factors that determine the spin
and brightness temperature (the observable in these experiments) and
show a number of results obtained from numerical simulations. In
\S~\ref{sec:observation} the expected foregrounds, the instrument
sensitivity and the predicted power spectrum measurement and the
skewness are discussed. The paper ends with conclusions.

\section{Current observational constraints}
\label{sec:constraints}

\sn Currently, there are a few strong observational constraints on the
EoR. The CMB temperature and polarization data obtained by the WMAP
satellite allow measurement of the total Thomson scattering of the
primordial CMB photons off intervening free electrons produced by the
EoR along the line of sight \citep{page07, spergel07,
komatsu10}. They show that the CMB radiation has, on its way to us, only been damped by
$\sim 9\%$, indicating that the Universe was mostly neutral for 400
million years and then ionized. However, the Thomson scattering
measurement is an integral constraint telling us little about the
sources of reionization, its duration or how it propagated to fill the
whole Universe.

\sn The second strong constraint comes from specific features in the
spectra of distant quasars, known as the Lyman-alpha forest. These
features, which are due to neutral hydrogen, indicate two important
facts about reionization: 1) hydrogen in the recent Universe is highly
ionized with neutral fraction of $\sim 10^{-4}$; 2) at redshift 6.5,
i.e., about 900 million years after the Big Bang, the neutral fraction
of hydrogen suddenly increases demarcating the end of the reionization
process \citep{fan03, fan06}. Despite these data providing strong
constraints on the ionization state of the Universe at redshifts below
6.5, they say very little about the reionization process itself
\citep[although see][]{bolton07}.

\sn A third, albeit weaker, constraint comes from measuring the
temperature of the intergalactic medium at the redshift range 4-6.
The widths of the hydrogen Lyman-$\alpha$ absorption lines can be used
to determine the IGM temperature.  This property has been used by a
number of authors who have measured the IGM temperature at z$\approx
4$ \citep{schaye00,theuns02b}.  The temperature of the diffuse IGM
at these redshifts depends on its reionization history because the
thermal timescales are long. \cite{theuns02b} and \cite{hui03}, have used these measurement at redshift
about 4 to extrapolate back in time where they argue that, for a
single-phase reionization scenarios, hydrogen reionization occurred
below redshift z=9. Recently, \cite{bolton10} have
measured the IGM temperature at even higher redshift (z$\approx 6$)
and employed similar arguments to conclude that reionization happened
at redshift lower than 9.  As said, this constraint is somewhat weaker
than the previous two as it requires assumptions about the heating and
cooling rates from the redshift at which the temperature is measured
all the way back to the reionization redshift.

\sn A whole slew of possible other constraints currently discussed in the
literature are either very controversial, very weak or, as is often
the case, both. Most are very interesting and exciting; but can only
be investigated reliably with a new generation of instruments that are
expected to come online within a decade.

\begin{figure} \centering
   \includegraphics[width=8.5cm,clip]{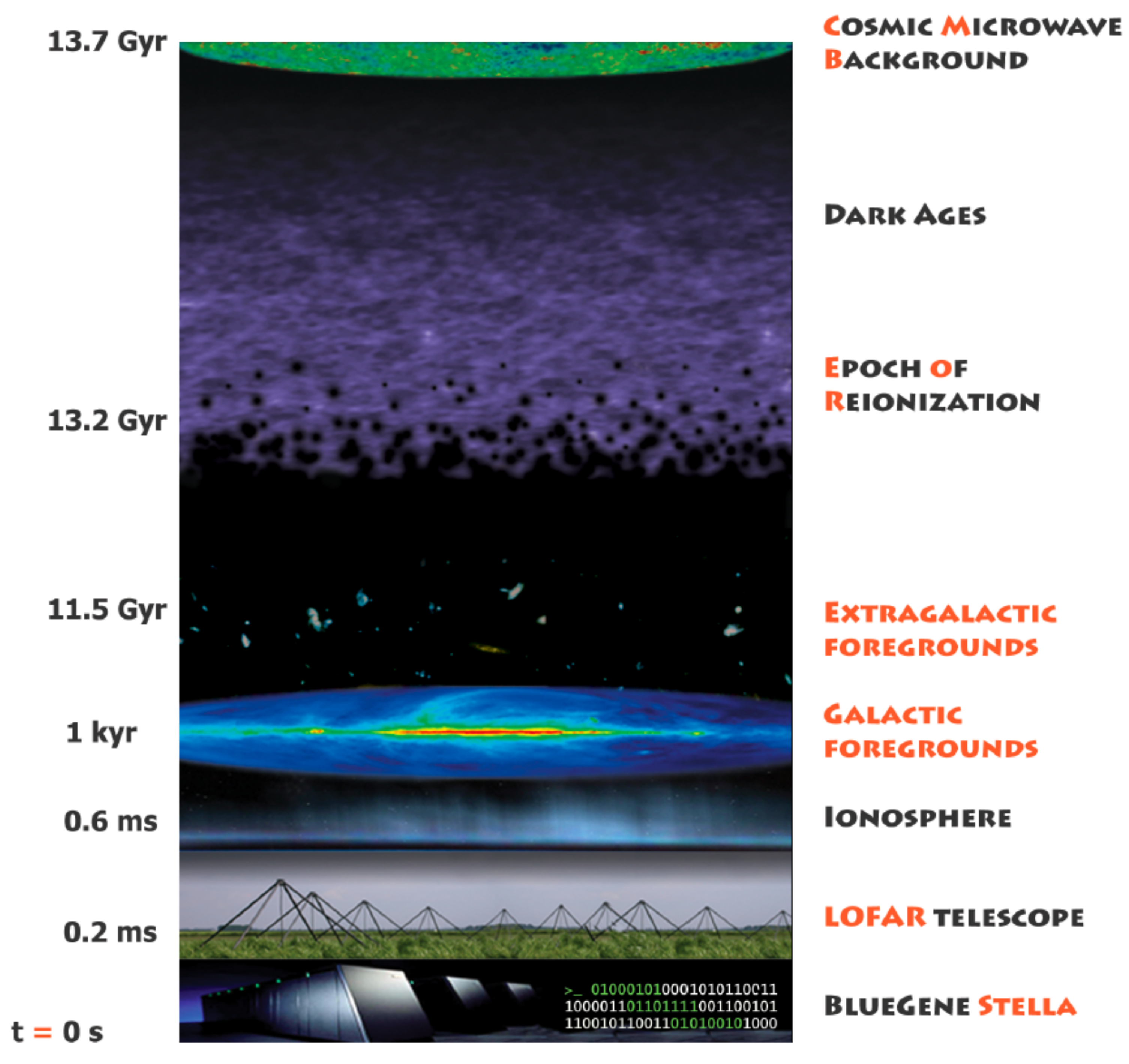}
   \caption{\footnotesize This figure shows a sketch of the likely
development of the EoR.  About 500,000 years after the Big Bang ($z
\sim 1000$) hydrogen recombined and remained neutral for a few hundred
million years (the \textit{Dark Ages}). At a redshift, $z \sim 15$,
the first stars, galaxies and quasars began to form, heating and
reionizing the hydrogen gas. The neutral IGM can be observed with
LOFAR through its redshifted 21cm spin-flip transition up to redshift
11.5. However, many atmospheric, galactic and extra-galactic
contaminants corrupt the 21~cm signal.  }
   \label{fig:lofareor}
    \end{figure}

\sn To summarize, from the current observational constraints and recent
numerical and theoretical models of structure formation a very simple
picture of the Universe's reionization history has emerged.  A cartoon
of the various phases and the objects featuring in this simple picture
is shown in Figure~\ref{fig:lofareor}.

\section{Astrophysical Ionization Sources}
\label{sec:sources}

 \sn The nature of the first ionizing sources is unclear. In general
two types of astrophysical sources are considered\footnote{Other
ionization sources, e.g., decaying dark matter particles, are possible
but less likely} Population III stars or miniquasars, both of which
represent hypothetical but plausible populations of the first objects
in the universe, and that are significant sources of ionizing photons.

\sn  \textbf{\textit{First stars:}} The impact of stellar sources have been
studied by many authors \citep[e.g.,][]{cen03, ciardi03b, haiman03,
sokasian03, wyithe03}.  These studies find in general that in order
to provide enough ionizing flux at or before $z=15,$ for the usual
scale-invariant primordial density fluctuation power spectrum, one
needs Population III stars, which provide about 20 times more ionizing
photons per baryon than Population II \citep{schaerer02, bromm01}, or
an IMF that initially is dominated by high mass stars. This is in
agreement with recent numerical simulations of the formation of first
stars from primordial molecular clouds that suggest that the first
metal-free stars were predominantly very massive, $m_*\lsim 100
M_\odot$ \citep{abel00, bromm02}.
 
\sn  \textbf{\textit{Mini-quasars:}} Miniquasars have also been considered as a
significant ionising source \citep[e.g.,][]{kuhlen06, oh01, madau04,
ricotti04a, ricotti04b, thomas08, zaroubi05, zaroubi07}.  In view of the
correlation between central black hole mass and spheroid velocity
dispersion \citep{ferrarese02, gebhardt00}, this correlation
demonstrates that seed black holes must have been present before
spheroid formation.  Theory and observations suggests that the seeds
from which the super massive black holes amounted to at least $1000
\rm M_\odot$ and were in place before $z\sim 10$ \citep{fan03, fan06,
madau01, silk98}.

\sn \textbf{\textit{Decaying Dark Matter particles:}} In the literature, the influence
of dark matter particles annihilation or decay on reionization is considered.
The consensus here is that, given the constraints that
the CMB impose on the value of Thomson optical depth, these 
process are unlikely to substantially contribute to the Universe's
ionization. Instead many authors consider whether the
influence of annihilation/decay of dark matter particles on 
reionization could be used, despite its small amplitude, to constraint 
the properties of these particles.

\section{The 21~cm Spectral Line as a Probe of the EoR}
\label{sec:signal}

\sn In recent years it has become clear that the 21~cm line can be
used to probe the neutral IGM prior to and during reionization.  This
hyperfine transition line of atomic hydrogen (in the ground state)
arises due to interactions between and the electron and proton spins
\citep{hogan79, scott90, madau97}.  The excited triplet state is a
state in which the spins are parallel whereas the spins at the lower
(singlet) state are antiparallel. The 21~cm line is a forbidden line
with a probability of $2.9\times 10^{-15} s^{-1} $ corresponding to a
life time of about $10^7$ years.  Despite its low decay rate the 21~cm
transition line is one of the most important astrophysical probes,
simply due to the vast amounts of hydrogen in the Universe
\citep{ewen51, hulst45, muller51}

\subsection{The 21~cm Spin Temperature}


\sn The intensity of the 21~cm radiation is controlled by one
parameter, the so called spin temperature, $T_{spin}$. This temperature is 
defined through
the equation, $n_1 / n_0 = 3 \exp (- T_{\ast} / T_{spin})$, where
$n_1$ and $n_0$ are the number densities of electrons in the triplet
and singlet states of the hyperfine levels, and $T_{\ast} = 0.0681\
\mathrm{K}$ is the temperature corresponding to the 21~cm wavelength.

\sn In his seminal papers George Field \citep{field58, field59} used
the quasi-static approximation to calculate the spin temperature,
$T_{spin},$ as a weighted average of the CMB temperature, $T_{CMB}$,
the gas kinetic temperature, $T_{kin}$, and the temperature related to
the existence of ambient Lyman-$\alpha$ photons, $T_{\alpha}$
\citep{wouthuysen52, field59}. For almost all interesting cases one
can safely assume that $T_{kin}=T_{\alpha}$ \citep{field58,
furlanetto06, madau97}. The spin temperature is given by:
\begin{equation} T_{spin} = \frac{T_{CMB} + y_{kin} T_{kin} +
y_{\alpha} T_{kin}}{1 + y_{kin} + y_{\alpha}}, \label{eq:tspin}
\end{equation} where $y_{kin}$ and $y_{\alpha}$ are the kinetic and
Lyman-$\alpha$ coupling terms, respectively. The kinetic coupling term
is due to collisional excitations of the 21~cm transitions. The
Lyman-$\alpha$ coupling term is due to the so called Lyman-$\alpha$
pumping mechanism -- also known as the Wouthyusen-Field effect --
which is produced by photo-exciting the hydrogen atoms to their Lyman
transitions \citep{field58, field59, wouthuysen52}.  Finally, it is
important to note that for the 21~cm radiation to be observed it has
to attain a different temperature than that of the CMB background
\citep{field58, field59, hogan79, wouthuysen52}.

\sn Since the decoupling mechanisms can influence the spin temperature
differently, it is important to explore the decoupling issue for
various types of ionization sources. For instance, stars decouple the spin
temperature mainly through radiative Lyman$-\alpha$ pumping whereas
mini-quasars decouple it through a combination of collisional
Lyman$-\alpha$ pumping and heating \citep{chuzhoy06, zaroubi07}, both
produced by energetic secondary electrons ejected due to the miniqso
x-ray photons \citep{shull85}. The difference between the spin
temperature decoupling patterns between the two will eventually help
disentangle the nature of the first ionization sources \citep{thomas08}.

\begin{figure} \centering
   \includegraphics[width=8.5cm,clip]{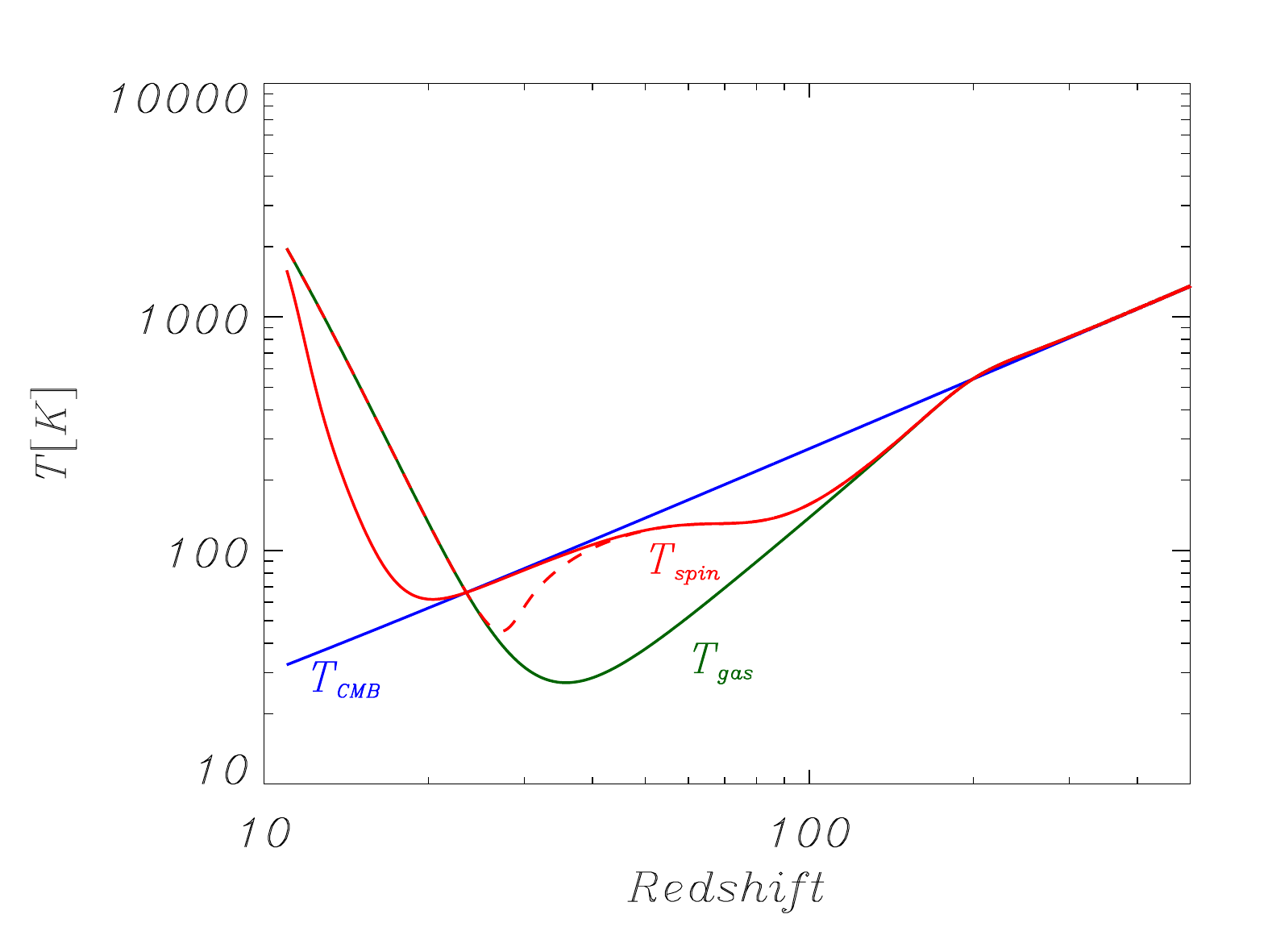}
   \caption{\footnotesize The global evolution of the CMB (blue line), gas
(green line) and spin (red solid line and red dashed line) temperatures as a function
of redshift. The CMB temperature evolves steadily as 1+z whereas the
gas and spin temperatures evolve in a more complicated manner (see
text for detail).  }
            \label{fig:TspinHist}
           
    \end{figure}

\sn Figure~\ref{fig:TspinHist} shows the expected global evolution of
the spin temperature as a function of redshift.  The blue solid line
represents $T_{CMB}$, which drops as $1+z$. The green line shows the gas
temperature as a function of redshift. At $z\gsim 200$ the gas
temperature is still coupled to the CMB due to Compton scattering off
residual electrons leftover from the recombination era. At redshift
$\sim 200$, however, the gas decouples from the CMB and starts
adiabatically cooling as a function of $(1+z)^2$ until the first
objects start forming and heating up the gas at redshift below 30.
The spin temperature (shown by the red lines) has a somewhat more
complicated behavior, at $z>30$ it is coupled to the gas temperature
due to collisional coupling caused by a residual electrons leftover
from recombination. At $z\approx100$
the efficiency of collisional coupling to the gas drops due to the Hubble expansion. At this stage,
the spin temperature starts veering towards $T_{CMB}$ until it is
completely dominated by it.  At lower redshifts the first
astrophysical objects that heat and ionize the IGM couple
$T_{spin}$ to the gas. Here, broadly speaking, there are two possible
histories, one in which $T_{spin}$ couples to the gas as it heats up
where it obtained temperature greater than $T_{CMB}$ (red solid
line). In the other possible evolution the spin temperature couples to
the gas much before the kinetic temperature exceeds that of the CMB
(red dashed line) \citep{thomas10, baek09}.  In the former case the
21~cm radiation, after decoupling from the CMB at $z\lsim 30$, is seen
only in emission, whereas in the latter case it is seen initially
in absorption and only at later stages in
emission.

\subsection{The Brightness Temperature}


\sn In radio astronomy, where the Rayleigh-Jeans law is usually
applicable, the radiation intensity, $I (\nu)$ is expressed in terms
of the brightness temperature, so that
\begin{equation} I (\nu) = \frac{2 \nu^2}{c^2} k_{B} T_b,
\end{equation} where $\nu$ is the radiation frequency, $c$ is the
speed of light and $k_{B}$ is Boltzmann's constant \citep{rybicki86}.
This in turn can only be detected differentially as a deviation from
$T_{CMB}$. The predicted
differential brightness temperature deviation from the cosmic
microwave background radiation is given by \cite{field58, field59,
ciardi03a},
\begin{eqnarray} \delta T_b = 28 \mathrm{{mK}} & \left( 1 + \delta
\right) & x_{HI} \left( 1 - \frac{T_{CMB}}{T_{spin}} \right)\times
\nonumber\\ & & \times \left( \frac{\Omega_b h^2}{0.0223} \right)
\sqrt{\left( \frac{1 + z}{10} \right) \left( \frac{0.24}{\Omega_m}
\right)},
\label{eq:dTb}
\end{eqnarray} where $h$ is the Hubble constant in units of $100~
\mathrm{{km} \, s^{- 1} {Mpc}^{- 1}}$, $\delta$ is the mass density
contrast, $x_{HI}$ is the neutral fraction, and $\Omega_m$ and
$\Omega_b$ are the mass and baryon densities in units of the critical
density. Note that the three quantities, $\delta$, $x_{HI}$ and
$T_{spin}$, are all functions of 3D position.

\sn One also should add to this equation the contribution of
redshift distortion. In the linear regime of gravitational instability
this component is simple to add \citep{kaiser87}. However, in the
quasi-linear regime this term is not well understood and should be
explored in more detail.

\begin{figure*} \centering
    \includegraphics[width=18cm,clip]{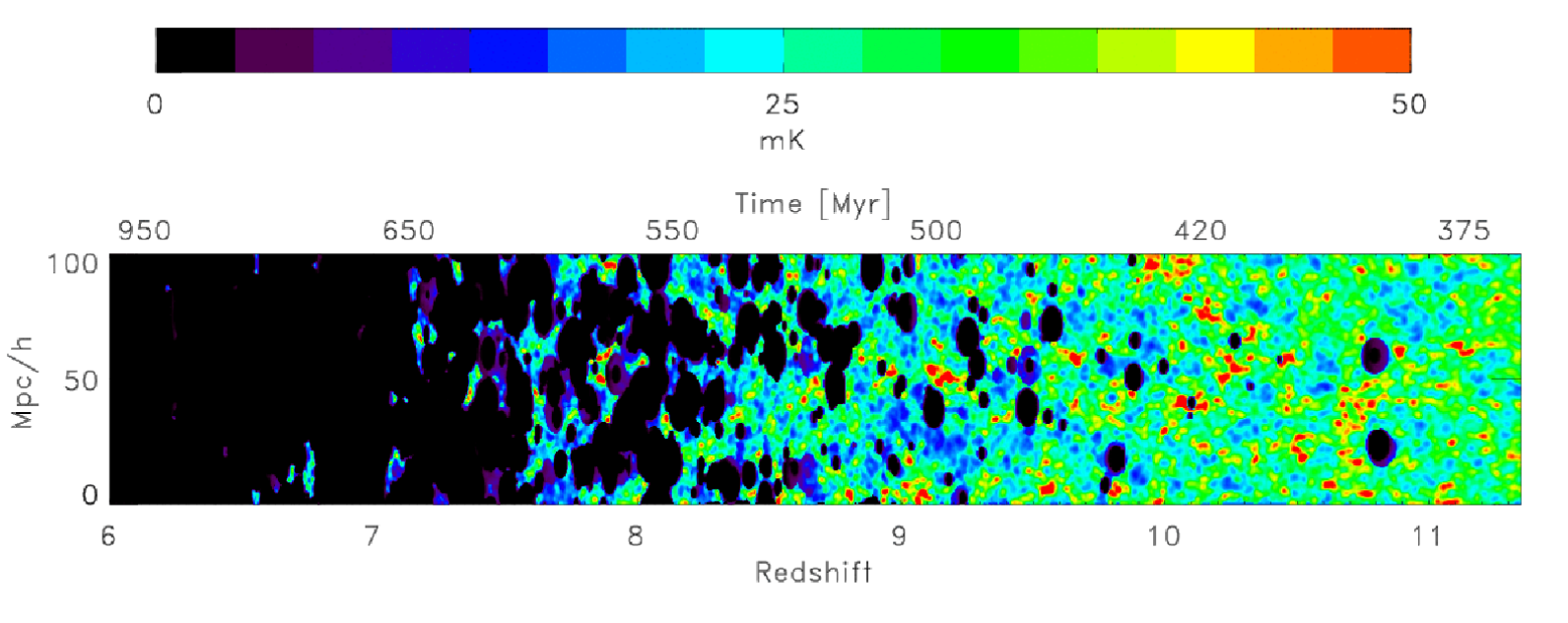}
      \caption{\footnotesize Position-redshift slices through the
image-frequency volume of the reionization simulations of \cite{thomas09}.
         \label{fig:EoRslice} }
   \end{figure*}

\sn Equation~\ref{eq:dTb} gives the cosmological signal we are after
and clearly shows the complication and, at the same time, the wealth
of information that could be extracted from the EoR signal. For
example, according to recent simulations \citep[e.g.,][]{iliev08,
thomas09} the ionized fraction of the IGM at $z\approx 11$ is very
small. If the spin temperature is much larger than $T_{CMB}$ the
brightness temperature is proportional to the cosmological density and
can provide a clean probe of the primordial density field. However, at
later stages one can probe the evolution of the ionized fraction as a
function of redshift which provides information about the ionization
sources and their clustering properties and so
on. Figure~\ref{fig:EoRslice} shows a typical distribution of the
differential brightness temperature as predicted by the recent \cite{thomas09} simulations.

\sn One should also note that in most simulations the spin temperature
is assumed to be much larger than the CMB temperature, namely
term $\left(1-{T_{CMB}}/{T_{spin}}\right)$ in eq.~\ref{eq:dTb} is $1$.  As
figure~\ref{fig:TspinHist} shows, this is a good assumption at the
later stages of reionization, however, it is not valid at the early
stages. Modeling this effect is somewhat complex and requires radiative
transfer codes that capture the Lyman-$\alpha$ line formation and
multifrequency effects, specially those coming from energetic photons
\citep{baek09, thomas10}.

\section{The redshifted 21~cm Observation}
\label{sec:observation}

\sn The current generation of radio telescopes
(LOFAR\footnote{http://www.lofar.org \&
http://www.astro.rug.nl/\~\,LofarEoR},
MWA\footnote{http://www.mwatelescope.org/}, GMRT, and
21CMA\footnote{http://21cma.bao.ac.cn/}) will capture the lower
redshift part of the $\delta T_b$  evolution ($z\lsim 12$).  LOFAR, for 
example have two observational bands, high
band and low band. The high band array is expected to be sensitive
enough for measuring the redshifted 21~cm radiation coming from the
neutral IGM within the redshift range of z=11.4 (115 MHz) to z=6 (203
MHz), with a resolution of 3-4 arcminutes and a typical field of view
of $\sim 120$ square degrees (with 5 beams) and a sensitivity of the
order of 80 mK per resolution element and 1 MHz frequency bandwidth.  At frequencies below
the FM band, probed by the low band array, the LOFAR
sensitivity drops significantly and the sky noise increase
dramatically that detection of HI signals at these frequencies is
beyond reach with LOFAR \citep{harker10,jelic08,panos09}.  The main
and most challenging task of the LOFAR EoR project is to extract the
cosmological signal from the data and interpret it.

\sn In the future SKA can significantly improve on the current
instruments in two major ways.  Firstly, it will have at least an
order of magnitude higher signal-to-noise which will allow much better
statistical detection of the EoR and give us access to the Universe's
\textit{\textit{Dark Ages}} which corresponds to much higher redshifts
($z\approx 35$) hence providing crucial information about cosmology
which non of the current telescope hopes to have. Secondly, SKA will
have a resolution better by a factor of few at least relative to the
current telescopes. These two advantages will not only improve on the
understanding we gain with current telescopes but give the opportunity
to address a host of fundamental issues that current telescope
will not be able to address at all. Here I give a few examples: 1- Due to
the limited resolution and poor signal-to-noise, the nature of the
ionizing sources is expected to remain poorly constrained;
2- the mixing between the astrophysical effects and the cosmological
evolution is severe during the EoR but much less so during the
\textit{Dark Ages} an epoch beyond the reach of the current
generation of telescope but within SKA reach; 3- at redshifts
larger than 30 the 21~cm could potentially provide very strong
constraints, much more so than the CMB, on the primordial
non-gaussianity of the cosmological density field which is essential
in order to distinguish between theories of the very early Universe
(e.g., distinguish between different inflationary models).

\sn In section~\ref{sec:signal} we discussed the cosmological 21~cm
signal and showed that it is expected to be of the order of $\approx
10$mK, However, the detectable signal in the frequency range that
corresponds the epoch of reionization is composed of a number of
components each with its own physical origin and statistical
properties.  These components are : 1- the 21~cm signal coming from
the high redshift Universe. 2- galactic and extra-galactic
foregrounds, 3- ionospheric influences, 4- telescope response effects
and 5- noise (see Figure~\ref{fig:lofareor}).  Obviously, the challenge
of the upcoming experiments is to distill the cosmological signal out
of this complicated mixture of influences. This will crucially depend
on the ability to calibrate the data very accurately so as to correct for the ever changing
ionospheric effects and variation of the instruments response with
time. In this section I present the various non cosmological
components and briefly discuss their properties.

\subsection{Station configuration and uv coverage}

\sn The low frequency arrays must be configured so that they have a very 
good uv-coverage. This is crucial to the calibration effort of the data where a
 filled uv plane is important for obtaining precise Local \citep{nijboer06} and Global
\citep{smirnov04} Sky models (LSM/GSM; i.e.  catalogues of the
brightest, mostly compact, sources in and outside of the beam,
i.e. local versus global).  It is also crucial for the ability to
accurately fit for the foregrounds \citep{harker09b, jelic08} and to the measurement of the EoR 
signal power spectrum \citep{bowman06, harker10, hobson02, santos05}

\sn The uv coverage of an interferometric array depends on the layout
of the stations (interferometric elements), their number and size as
well as on the integration time, especially, when the number of stations
is not large enough to have a good instantaneous coverage.

\sn For a given total collecting area one can achieve a better uv
coverage by having smaller elements (stations). For example LOFAR has
chosen to have large stations resulting in about $\approx 10^3$
baselines in the core area.  Such a small number of baselines needs
about 5-6 hours of integration time per field in order to fill the uv
plane (using the Earth's rotation). In comparison, MWA which has
roughly 1/2 the total collecting area of LOFAR chose to have smaller
stations with about $\approx 10^5$ baselines resulting in an almost
instantaneous full uv coverage.

\sn The decision on which strategy to follow has to do with a number
of considerations that include the ability to store the raw
visibilities, hence, allowing for a better calibration, an acceptable
noise level for both the foreground extraction needs as well the power
spectrum measurement (see the following sections; \S~\ref{sec:noise}
and \S~\ref{sec:foregrounds}).  A compromise between these issues as well
the use of the telescopes for other projects is what drives the final
lay out of the antennas.

\subsection{Noise Issues}
\label{sec:noise}

\sn The thermal noise level for a given visibility, i.e., uv point,
is,

\begin{equation} {\Delta V}(u,v) \approx \frac{2 k_{B}
T_{sys}}{\epsilon dA\sqrt{B t}}\, ,
\label{eq:vnoise}
\end{equation} where $T_{sys}$ is the system temperature, $\epsilon$
is the efficiency, $dA$ is the station area, $B$ is the bandwidth and
$t$ is the observation time \citep[see e.g.,][]{morales05}. This
expression is simple to understand, the more one observes, either in
terms of integration time, frequency band width or station collecting
area the less uncertainty one has. Obviously, if the signal we are
after is well localized in either time, space or frequency the
relevant noise calculation should take that into account.

\begin{figure} \centering
    \includegraphics[width=8.5cm]{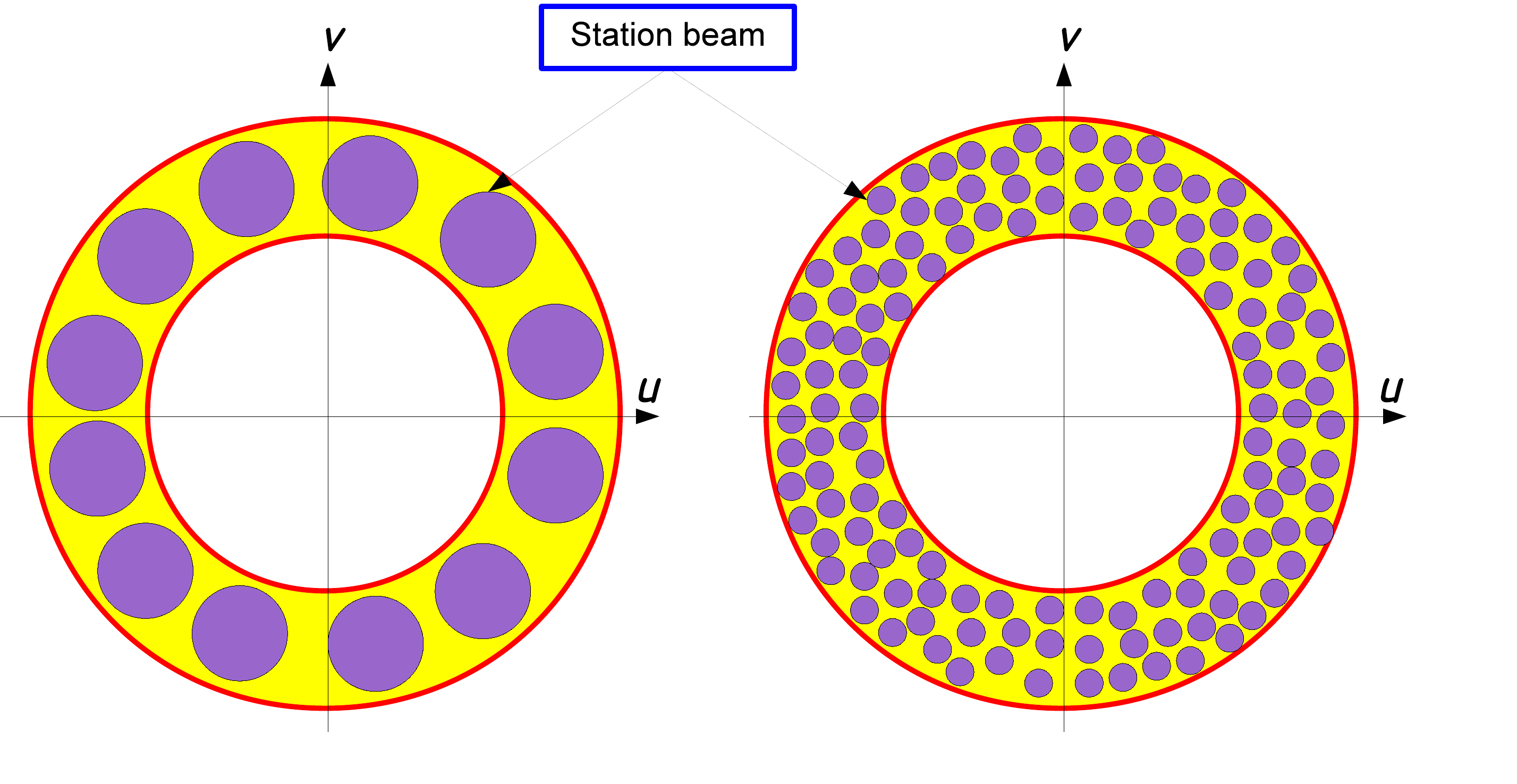}
      \caption{\footnotesize This figure shows how two different
experiments might sample an annulus in uv. The size of uv point is
given by the station (interferometric element) size, larger station
(left panel) has a larger footprint relative to the smaller station
case (right panel) in uv plane; the footprint is shown by the purple
circles. Even though the sampled area in the two cases might be the
same, the fact that smaller stations sample the annulus more results
in an increased accuracy in their estimation of the power spectrum.  }
          \label{fig:UVannulus}
   \end{figure}

\sn In order to calculate the noise in the 3D power spectrum, the main
quantity we are after, one should remember that the frequency
direction in the observed datacube is proportional to the redshift
which in turn can be easily translated to distance whereas the u and v
coordinates are in effect Fourier space coordinates. Therefore, to
calculate the power spectrum first first shoud Fourier transform the
data cube along the frequency direction. Following \cite{morales05} I will call the new Fourier space coordinate $\eta$
(with $d\eta$ resolution) which together with u and v define the
Fourier space vector $\vec{u}=\{u,v,\eta\}$.  From this, one can
calculate the noise contribution to the power spectrum at a given
$\vert\vec{u}\vert$,

\begin{equation} P_{noise}(\vert \mathbf{u}\vert) \approx 2
N_{beam}^{-1} N_{cell}^{-1/2} \left(\frac{2 k_{B} T_{sys}}{\epsilon dA
d\eta}\right)^2 \frac{1}{B n(\vert\vec{u}\vert) t} \, ,
\label{eq:PSnoise}
\end{equation} where $N_{beam}$ is the number of simultaneous beams
that could be measured, $N_{cell}$ is the number of independent
Fourier samplings per annulus and $n(\vert\vec{u}\vert)$ is the number
of baselines covering this annulus \citep{morales05}. Note that
$n(\vert\vec{u}\vert)$ is proportional to square number of stations,
hence, $n(\vert\vec{u}\vert) dA^2$ is proportional to the square of
the total collecting area of the array regardless of the station
size. This means that the noise power spectrum measurement does not
depend only on the total collecting area, band width and integration
time, it also depends the number of stations per annulus. This is easy
to understand as follows, the power in a certain Fourier space annulus
is given by the variance of the measured visibilities in the annulus
which carries uncertainty proportional to the inverse square root of
number of points. This point is demonstrated in
figure~\ref{fig:UVannulus}. As an example, MWA will have the advantage
of having many samplings of the uv plane (100 times more than LOFAR)
whereas LOFAR will have the advantage of simultaneous multi-beams and
somewhat larger collecting area.

\subsection{The Foregrounds}
\label{sec:foregrounds}

\sn The foregrounds in the frequency regime ($40-200$MHz) are very
bright and dominate the sky. In fact the amplitude of the foreground
contribution, $T_{sky}$, at $150 \mathrm{MHz}$ is about 4 orders of
magnitude larger than that of the expected signal. However, since we
are considering radio interferometers the important part of the
foregrounds is that of the fluctuations which reduces the ratio
between the them and the cosmological signal  to about 2-3 orders of magnitude, which is still a
formidable obstacle to surmount.


   \begin{figure} \centering
    \includegraphics[width=8.5cm,clip]{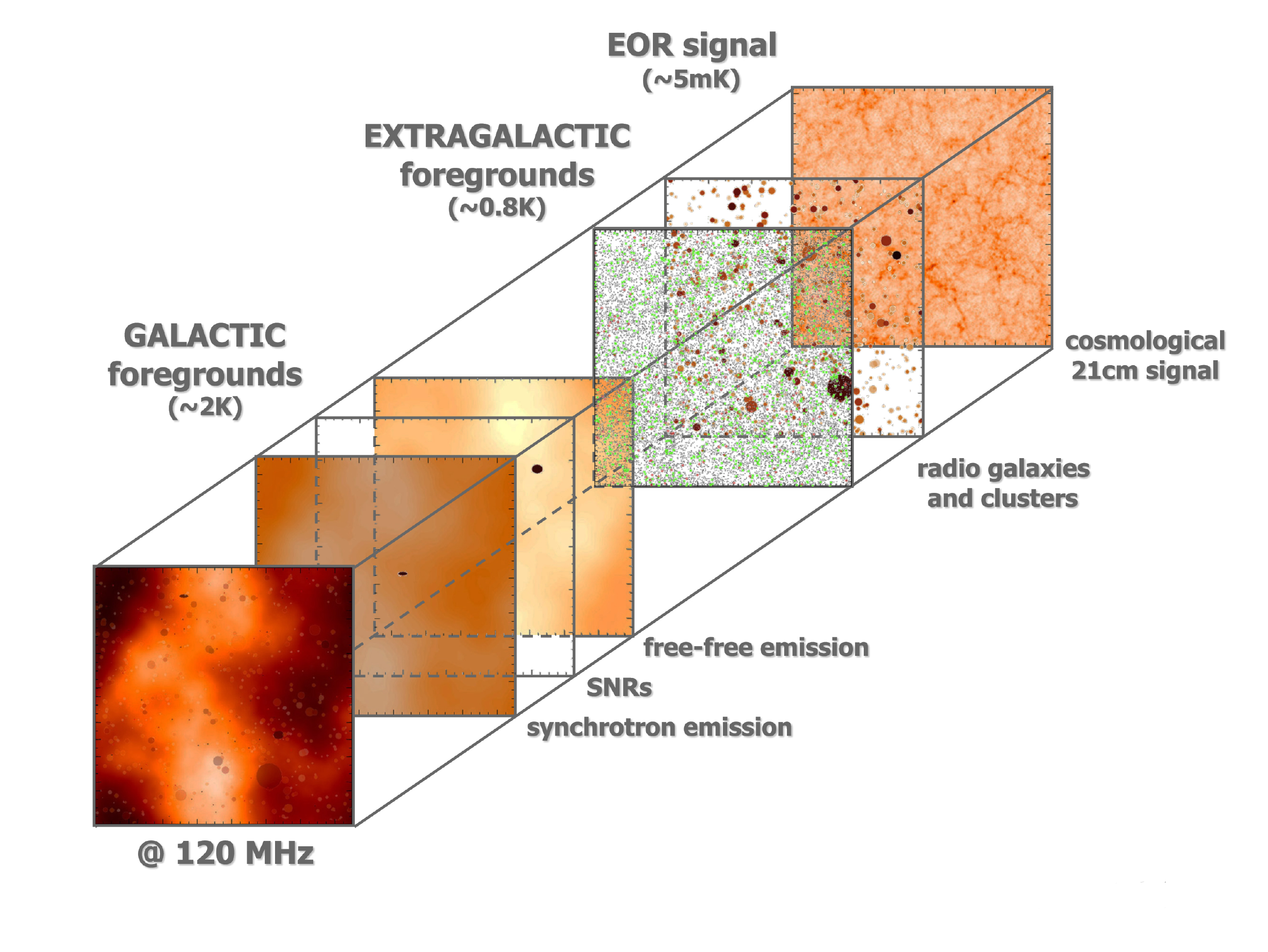}
      \caption{\footnotesize A figure showing the various cosmological
and galactic components that contribute to the measured signal at a
given frequency. The slices are color coded with different tales owing
to the vast difference between the range of brightness temperature in
each component, however the figure shows the rms of the galactic
foregrounds, extra galactic foregrounds and cosmological signal}
 \label{fig:FGslice}
  \end{figure}

 \sn The most prominent foreground is the synchrotron emission from
relativistic electrons in the Galaxy, this source of contamination
contributes about 75\% of the foregrounds. Other sources that
contribute to the foreground are radio galaxies, galaxy clusters,
resolved supernovae remnant, free-free emission provide 25\% of the
foreground contribution \citep[see][]{shaver99}.
Figure~\ref{fig:FGslice} shows simulated foreground contribution at
120 MHz taken into account all the foreground sources mentioned.

\begin{figure} \centering
    \includegraphics[width=8.5cm,clip]{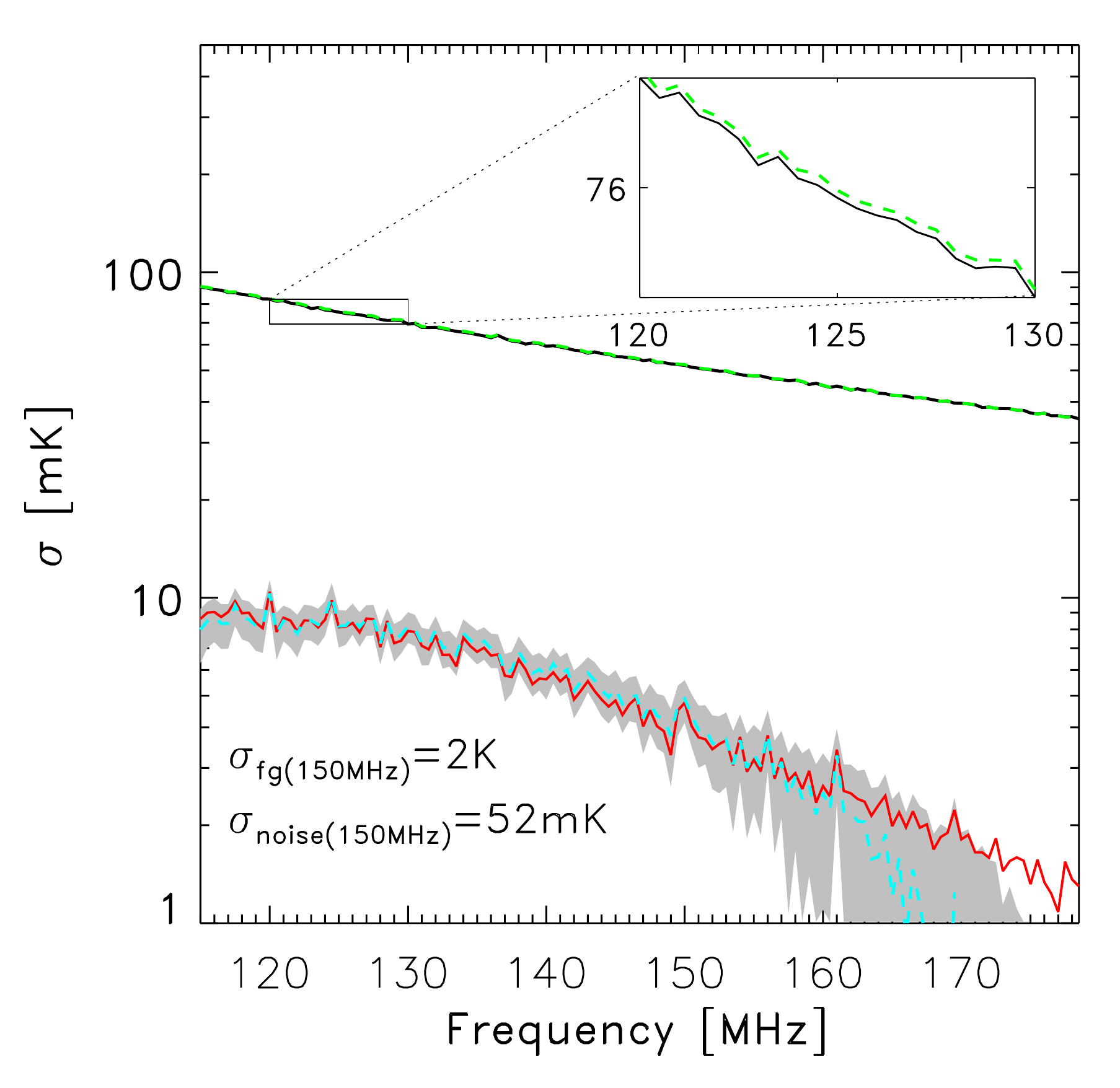}
    \caption{\footnotesize Statistical detection of the EoR signal
from the LOFAR-EoR data maps that include diffuse components of the
foregrounds and realistic instrumental noise ($\sigma_{noise}(150~{\rm
MHz})=52~{\rm mK}$). Black dashed line represents standard deviation
($\sigma$) of the noise as a function of frequency, cyan dashed line
$\sigma$ of the residuals after taking out smooth foregrounds
component and red solid line the $\sigma$ of original EoR signal. The
grey shaded surface represents the $90\%$ of detected EoR signals from
1000 independent realisations of the noise, where cyan dashed line is
mean of detected EoR signal. Note that the y-axis is in logarithmic
scale \citep{jelic08}.}
\label{fig:FGfit}
\end{figure}

 \sn As many studies have shown, the very smooth structure of the
foreground sources along the frequency direction will enable
disentangling their contribution from that of the cosmological signal.
The foregrounds are normally fitted by some 
procedure (e.g., polynomial fitting \citep{jelic08}, or more advanced non parametric
methods \citep{harker09b}) in order to recover the EoR cosmological
signal; Figure~\ref{fig:FGfit} shows how successful such a recovery
is.

\section{The Statistics of the observed cosmological signal}
\subsection{The 21~cm Power Spectrum}

\begin{figure} \centering
    \includegraphics[width=8.5cm]{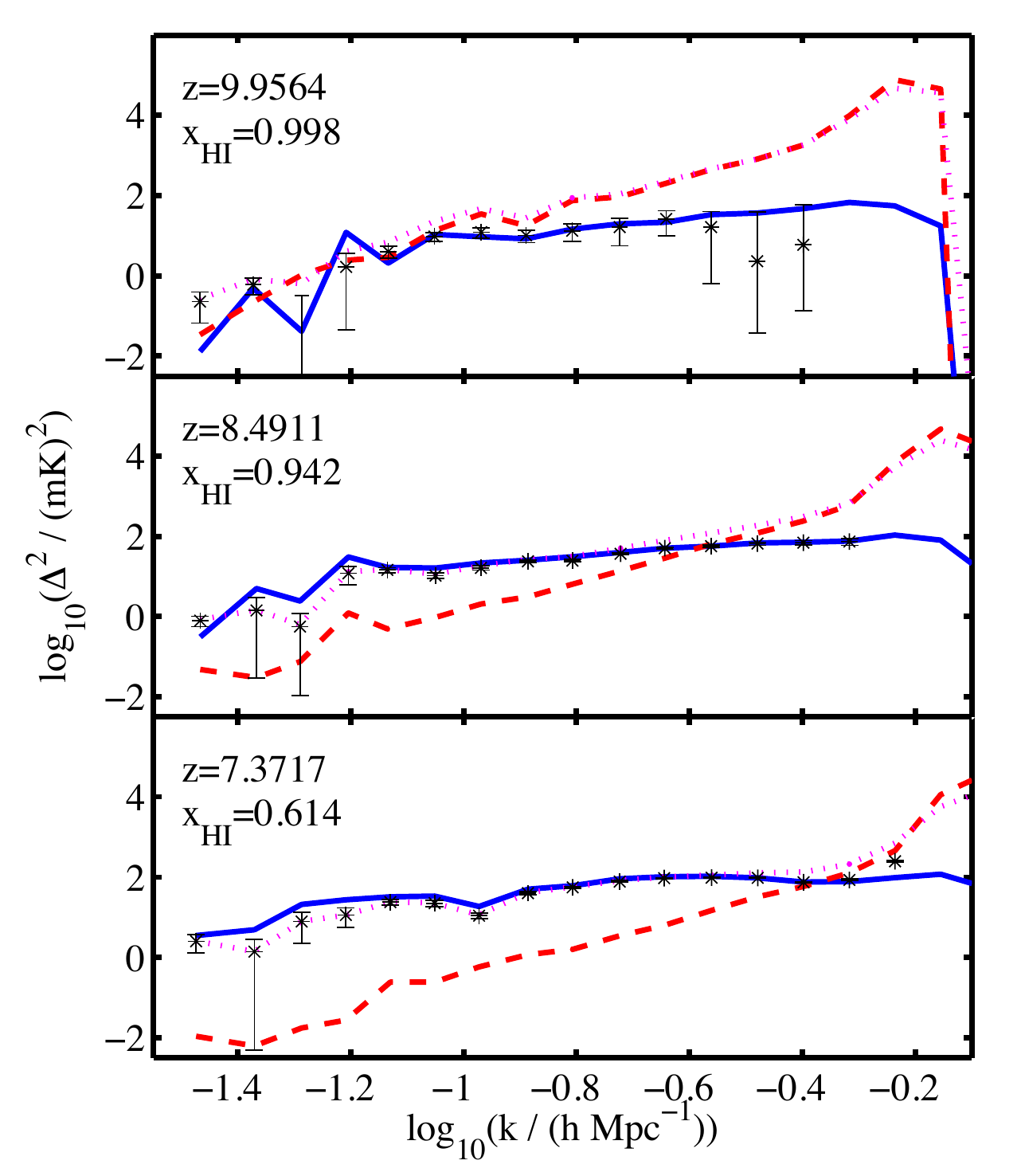}
      \caption{\footnotesize Power spectra of the cosmic signal (blue
solid line), the noise (red dashed line), the residuals (magenta
dotted line) and the extracted signal (black points with error bars)
at three different redshifts.for the case when the uv coverage is
frequency-dependent, we have 300~hours of observation per frequency
channel with a single station beam, and the foreground fitting is done
using Wp smoothing in Fourier space \citep{harker10}.  }
          \label{fig:PS}
   \end{figure}

\sn One of the main aims of the EoR projects is to measure the power
spectrum of variations in the intensity of redshifted 21-cm radiation
from the EoR.  As shown in Equation~\ref{eq:dTb} the power spectrum
depends on a number of astrophysical and cosmological quantities. The
sensitivity with which this power spectrum can be estimated depends on
the level of thermal noise (Eq.~\ref{eq:PSnoise}) and sample variance,
and also on the systematic errors arising from the extraction process,
in particular from the subtraction of foreground contamination. In the
LOFAR case, for example, we model the extraction process using
realistic simulations of the cosmological signal, the foregrounds and
noise. In doing so we estimate the sensitivity of the LOFAR EoR
experiment to the redshifted 21~cm power spectrum. Detection of
emission from the EoR should be possible within 360 hours of
observation with a single station beam. Integrating for longer, and
synthesizing multiple station beams within the primary (tile) beam,
then enables us to extract progressively more accurate estimates of
the power at a greater range of scales and redshifts 
\citep[see Figure~\ref{fig:PS}][]{harker10}.

\subsection{High order statistics: Skewness and Kurtosis}

\begin{figure} \centering 
    \includegraphics[width=8.5cm]{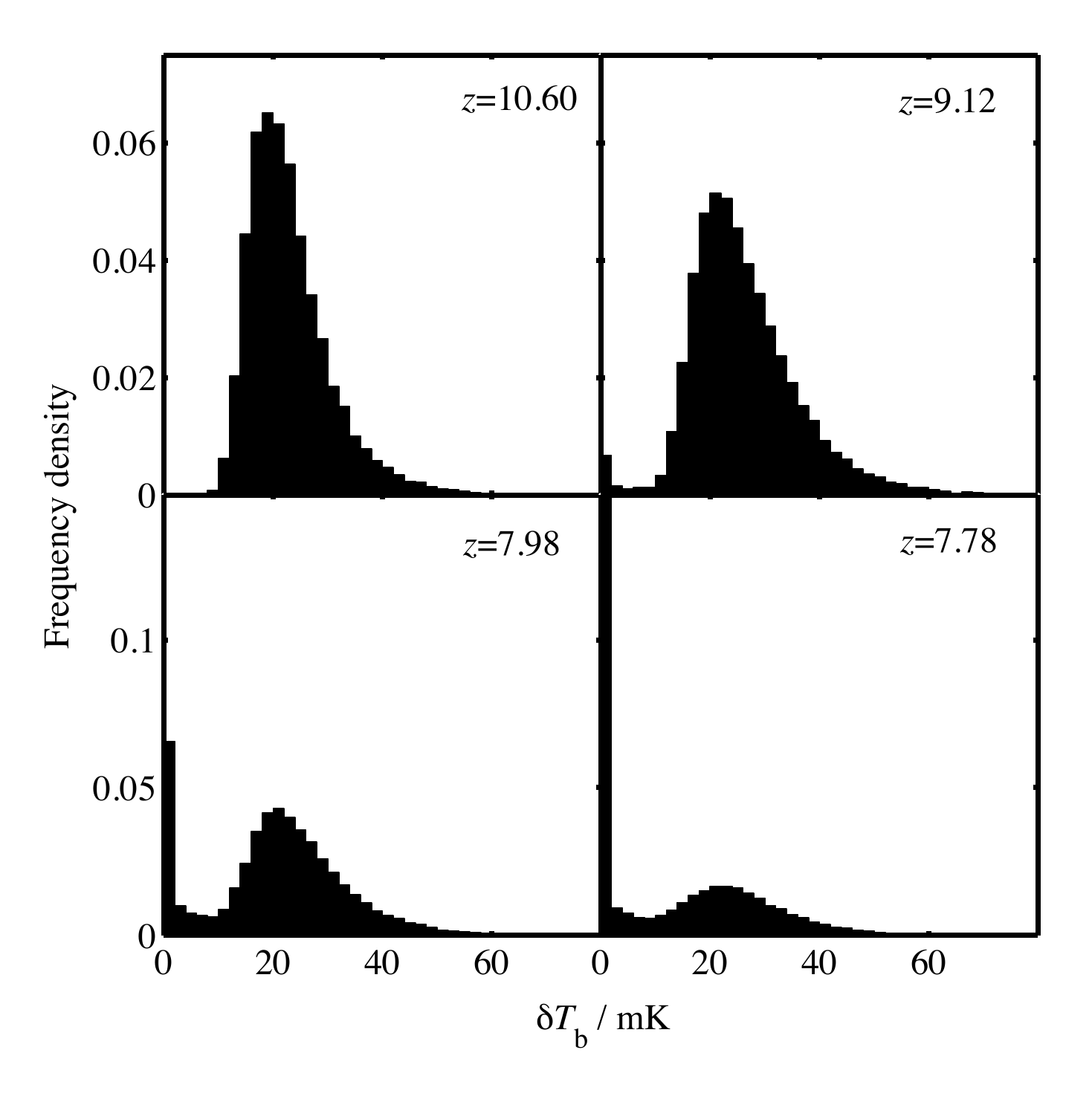}
\caption{\footnotesize The distribution of $\delta T_\mathrm{b}$ in a
certain cosmological simulation of reionization \citep{iliev08} at
four different redshifts, showing how the PDF evolves as reionization
proceeds. Note that the y-axis scale in the top two panels is
different from that in the bottom two panels. The delta-function at
$\delta T_\mathrm{b}=0$ grows throughout this period while the rest of
the distribution retains a similar shape. The bar for the first bin in
the bottom-right panel has been cut off: approximately 58 per cent of
points are in the first bin at $z=7.78$ \citep{harker09a}.}
    \label{fig:onepoint}
    \end{figure}

\begin{figure} \centering
    \includegraphics[width=8.5cm]{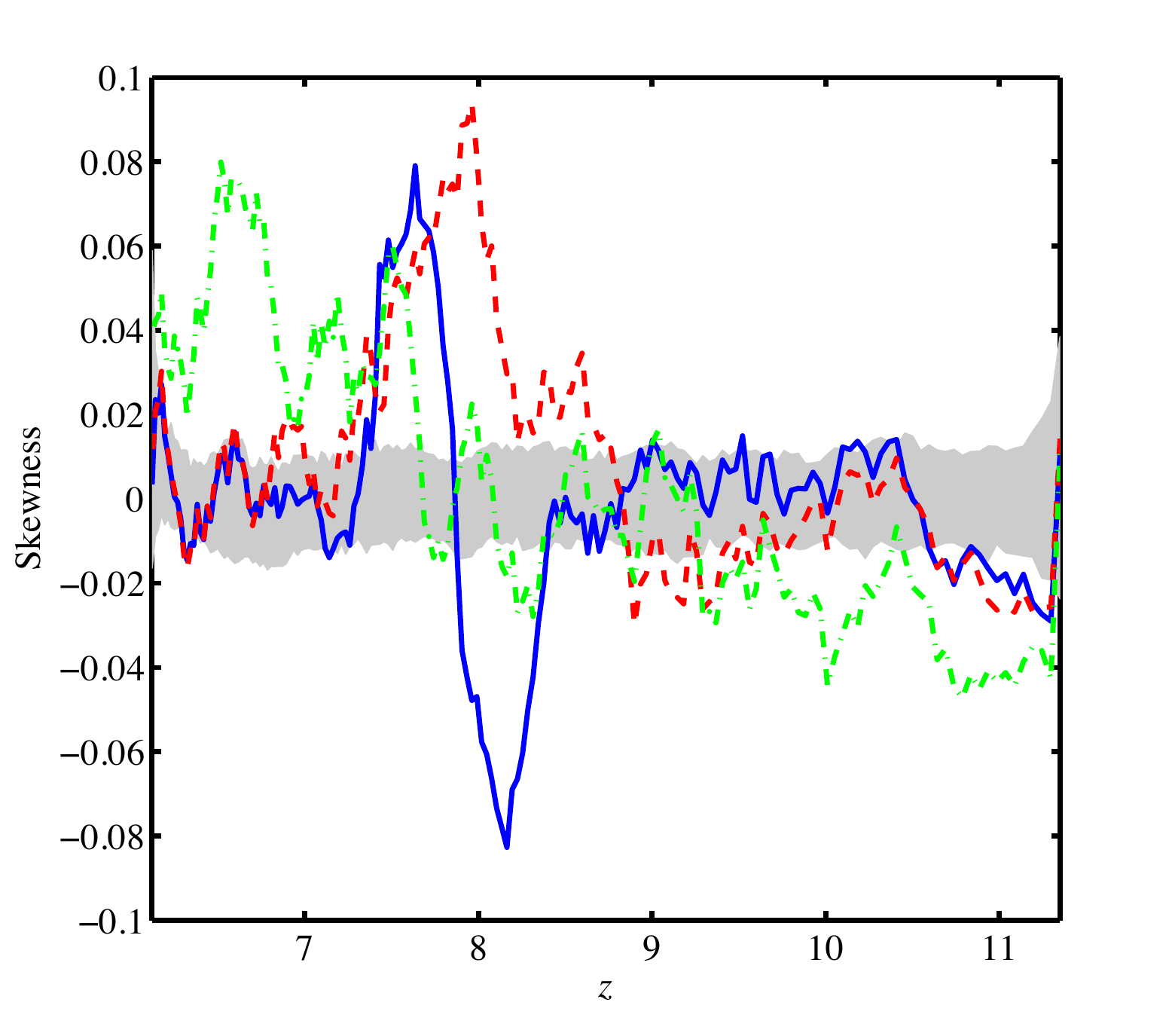}
      \caption{\footnotesize { Skewness of the fitting residuals from
data cubes with uncorrelated noise, but in which the residual image
has been denoised by smoothing at each frequency before calculating
the skewness. The three lines correspond to results from three
different simulations \citep{thomas09, iliev08}. Each line has been
smoothed with a moving average (boxcar) filter with a span of nine
points. The grey, shaded area shows the errors, estimated using 100
realizations of the noise \citep[see][]{harker09a} }}
          \label{fig:Skewness}
   \end{figure}

\sn Figure~\ref{fig:onepoint} showes the PDF of the brightness
temperature at four different redshifts; the PDF is clearly
nongaussian at all four cases. High order moments, like the skewness,
as a function of redshift could be a useful tool for signal extraction
in the presence of realistic overall levels of foregrounds and noise.
\cite{harker09a} have shown that the cosmological
signal, under generic assumptions, have a very characteristic pattern
in the skewness as a function of redshift
(Figure~\ref{fig:Skewness}). At sufficiently high redshifts the signal
is controlled by the cosmological density fluctuations which, in the
linear regime, are Gaussian. At lower redshifts, and as nonlinearity
kicks in, the signal starts getting a slightly positive skewness.  As
the ionization bubbles begin to show up the skewness starts veering
towards 0 until it crosses it to the negative side when the weight of
the ionized bubbles becomes more important than the high density
outliers --note high density outliers are likely to ionize first-- but
the distribution is still dominated by the density fluctuations. At
lower redshift the bubbles dominate the PDF and the neutral areas
become the outliers giving rise to a sharp positive peak to the
skewness. At redshift around 6 the instrument noise assumed to be
Gaussian dominates driving the skewness again towards zero.
Exploiting this characteristic behavior might allow us to pick up the
cosmological signal with this high order statistic.

\section{Conclusions}

\sn The imminent availability of observations of redshifted 21~cm
radiation from the Universe's \textit{Dark Ages} and the EoR will be one
of the most exciting development in the study of cosmology and galaxy
and structure formation in recent years. Currently, there are a number
of instruments that are designed to measure this radiation. In this
contribution I have argued that despite the many difficulties that face
such measurements they will provide a major breakthrough in our
understanding of this crucial epoch. In particular current radio
telescopes, such as LOFAR, will be able to provide us with the global
history of the EoR progression, the fluctuations power spectrum during
the EoR, etc., up to $z\approx 11$. These measurements will usher the
study of the high redshift Universe into a new era which will bridge,
at least in part, the large gap that currently exists in
observation between the very high redshift Universe ($z\approx 1100$)
as probed by the CMB and low redshift Universe ($z\lsim 6$).

\sn Although the current generation of telescopes have a great promise 
they will also have limitations.  For example they have neither the
resolution, the sensitivity nor the frequency coverage to address many
fundamental issues, like the nature of the first sources. Crucially,
they will not provide a lot of information about the \textit{Dark
Ages} which is only accessible through very low frequencies in the
range of $40-120$ ($z\approx 35-11$).

\sn Fortunately, in the future SKA can improve dramatically on the
current instruments in three major ways.  Firstly, it will have at
least an order of magnitude higher signal-to-noise which will allow
much better statistical detection of the EoR. Secondly, it will give
us access to the Universe's \textit{Dark Ages} which corresponds to
much higher redshifts ($z\sim30$) hence providing a truly pristine
probe of cosmology. Thirdly, SKA will have, at
least, a factor of few better resolution in comparison with the
current telescopes, thus better constraining the dominant ionization
sources. In summary these three advantages will not only improve on
the understanding we gain with current telescopes but give the
opportunity to address a host of fundamental issues that current
telescope will not be able to address.
 
\sn The next decade will be extremely exciting for studying the high
redshift Universe, especially as these radio telescope gradually come
online, starting with LOFAR and MWA. As they promise to resolve
many of the puzzles we have today pertaining to the formation and
evolution of the first object, cosmology and the physical process in
the high redshift intergalactic medium.

\begin{acknowledgements} Many of the results shown here have been
obtained in collaboration with the members of the LOFAR EoR project
whose contribution I would like to acknowledge.
\end{acknowledgements}

\end{document}